\documentclass[aps,prl,twocolumn,showpacs,superscriptaddress,groupedaddress,floatfix]{revtex4-1}
\usepackage{graphicx}
\usepackage{amsmath}
\usepackage{amssymb}
\usepackage{amsfonts}
\usepackage{bm}
\usepackage{comment}
\usepackage{hyperref}
\newcommand{\be}{\begin{equation}}
\newcommand{\ee}{\end{equation}}
\newcommand{\bn}{\begin{eqnarray}}
\newcommand{\en}{\end{eqnarray}}

\usepackage{color} % use colored text in latex

\usepackage{hyperref}
\hypersetup{
colorlinks=true,final=true,
        linkcolor=red,
        citecolor=blue,
        filecolor=blue,
        urlcolor=blue,
}
\begin{document}

\title{Quantum critical dynamics of a magnetic impurity in a semiconducting host}

\author{Nagamalleswararao Dasari$^{1}$}\email{nagamalleswararao.d@gmail.com}
\author{Swagata Acharya$^{2}$}
\author{A. Taraphder$^{2,3}$} 
\author{Juana Moreno$^{4,5}$}
\author{Mark Jarrell$^{4,5}$}
\author{N. S. Vidhyadhiraja$^{1}$}\email{raja@jncasr.ac.in}
\affiliation{$^{1}$Theoretical Sciences Unit, Jawaharlal Nehru Centre For
Advanced Scientific Research, Jakkur, Bangalore 560064, India.}
\affiliation{$^{2}$Department of Physics, Indian Institute of Technology 
Kharagpur, Kharagpur 721302, India.}
\affiliation{$^{3}$Centre for Theoretical Studies, Indian Institute of
Technology Kharagpur, Kharagpur 721302, India.}
\affiliation{$^{4}$Department of Physics $\&$ Astronomy, Louisiana State University, Baton Rouge, LA 70803, USA}
\affiliation{$^{5}$Center for Computation $\&$  Technology, Louisiana State University, Baton Rouge, Louisiana 70803, USA.}

\begin{abstract}
 We have investigated the finite temperature dynamics of the singlet to doublet 
continuous quantum phase transition in the gapped Anderson impurity model using 
hybridization expansion continuous time quantum Monte Carlo. Using the self-energy 
and the longitudinal static susceptibility, we obtain a phase diagram in the 
temperature-gap plane. The separatrix between the low temperature local
 moment phase and the high temperature generalized Fermi liquid phase of this 
phase diagram is shown to be the lower bound of the critical scaling region 
of the zero gap quantum critical point of interacting type. We have computed the nuclear magnetic spin-lattice relaxation rate,
the Knight shift and the Korringa ratio, which show strong deviations
for any non-zero gap from the corresponding quantities in the gapless Kondo screened impurity case. 
\end{abstract}

\maketitle

{\textit{Introduction.}}$\textbf{---}$The screening of a magnetic impurity by 
conduction electrons embodies the Kondo effect\cite{kondo,dot}; a quantum many body phenomenon
that arises in dilute metallic alloys and mesoscopic quantum dot systems,
and is well studied theoretically and experimentally.  A closely related problem, that 
remains to be fully understood, is that  of dilute magnetic impurities in a 
semiconducting bath, in particular their finite temperature dynamics. This problem is 
of direct relevance to conventional 
superconductors\cite{PhysRevLett.89.256801,PhysRevB.88.045101}, 
valence fluctuating insulators\cite{PhysRevB.92.161108,PhysRevLett.112.136401} and dilute magnetic 
semiconductors\cite{2108131,PhysRevB.76.195207}.  Theoretical 
investigations of this problem have focused on the gapped Anderson impurity model (GAIM),
which describes a correlated impurity coupled to a bath of conduction electrons 
whose density of states has a hard gap ($\delta$) at the Fermi-level.

The GAIM has been investigated using several analytical and numerically exact 
methods\cite{saso1,Takegahara,saso2,chen,moca,Pinto2012567,logan2,logan1}. Early results 
generated a debate about the ground state of the model,
namely about the minimum gap\cite{saso1,Takegahara,saso2} required to screen the impurity local moment
completely at $T=0$. A consensus has now been reached through numerical
renormalization group (NRG)\cite{chen,moca} and local moment approach (LMA)\cite{logan2,logan1} 
results that in the symmetric case, the boundary quantum phase transition from a Fermi liquid singlet
to a local moment doublet ground state occurs at a zero critical gap, i.e $\delta_c=0$.
In particular, the LMA yields a closed scaling form\cite{logan2} for the single particle 
spectral function. Further, the authors found that a Kondo resonance 
like feature survives\cite{logan2} only for $\delta \lesssim T_K$, which was confirmed by 
recent NRG calculations\cite{moca}. In a separate work, the authors proved a number of exact 
results using self-consistent perturbation theory to all orders\cite{logan1}, 
including e.g.\ that the ph-symmetric point of the GAIM is 
necessarily a non-Fermi liquid local moment phase, for all nonzero gaps.

Thus, a quantum critical point (QCP) at $\delta_c=0$ in the symmetric GAIM has
been established beyond doubt.
The study of critical properties and establishing the critical region for boundary 
quantum phase transitions, 
which can occur in variants of impurity Kondo models\cite{Fradkin,Ingersent,si,tsvelik,10.1080}, 
is one of the current 
research interests in condensed matter physics. In this context, the finite temperature
critical region and dynamics
of the GAIM have not been investigated. Presently, QCPs
are classified as interacting or non-interacting type based on the 
presence or absence respectively of $\omega/T$ scaling in the critical 
region\cite{PhysRevLett.107.076404,PhysRevB.91.035118,gegenwart2008quantum}. For 
the QCP in the GAIM, such an identification has not been carried out. 
In this work, we have studied the particle-hole symmetric case of the GAIM using the hybridization 
expansion version of the continuous time quantum Monte-Carlo (CTQMC)\cite{CTQMC1}.  
A crossover in single-particle dynamics from a low temperature local moment phase
to a high temperature generalized Fermi liquid phase is used to establish 
a phase diagram of the GAIM in the temperature vs. gap plane. The loci of such crossovers
in the phase diagram is shown, through an $\omega/T$ scaling of the dynamical susceptibility,
to be the lower bound of the critical scaling region of the $\delta_c=0$ {\em interacting type} QCP.
Finally, the magnetic relaxation rate, 
the Knight shift and the Korringa ratio are shown to exhibit highly
anomalous behaviour for all non-zero gaps.

\textit{Model and Formalism.}$\textbf{---}$The generic Anderson model that describes
 a quantum impurity coupled to a bath of conduction electrons is given by
\begin{equation*}
H=\sum_{k\sigma}\epsilon_k c^{\dagger}_{k\sigma} c^{\phantom \dagger}_{k\sigma}
+V\sum_{k\sigma}(c^{\dagger}_{k\sigma}d^{\phantom \dagger}_{\sigma}+h.c)+ 
\epsilon_d  n_d + U n_{d\uparrow} n_{d\downarrow}\,,
\end{equation*}
where $\epsilon_k$ is the host dispersion and V is the hybridization which 
couples the impurity to the bath. $\epsilon_d$ is the orbital energy for the non-dispersive 
local level and $U$ is the energy cost for double occupancy of the impurity. 
The bath Green's function in the Matsubara frequency space can be written as
%\begin{equation}
$\mathcal{G}^0(i\omega_n) = \left[i\omega_n -\epsilon_d - \Delta(i\omega_n)\right]^{-1}\,,$
%\end{equation}
where $\Delta(i\omega_n)$ is the hybridization function. For the GAIM, this
 is given by
\begin{equation}
\Delta(i\omega_n)=-\frac{iV^2}{D-\delta}\left[\tan^{-1}\left(\frac{D}{\omega_n}\right)
-\tan^{-1}\left(\frac{\delta}{\omega_n}\right)\right]\,,
\end{equation}
which corresponds to a flat density of states with half-bandwidth $D$ and a 
gap of 2$\delta$ at the Fermi level. We employ the hybridization expansion 
CTQMC\cite{CTQMC1} to measure the dynamical quantities such as single and two particle 
Green's functions. The hybridization expansion CTQMC method yields data on the 
Matsubara axis. The maximum 
entropy method\cite{Jarrell} is used subsequently to obtain the real frequency
dynamical spin susceptibility.

\begin{figure}[t!]
\centering
\includegraphics[angle=0,width=1.0\columnwidth]{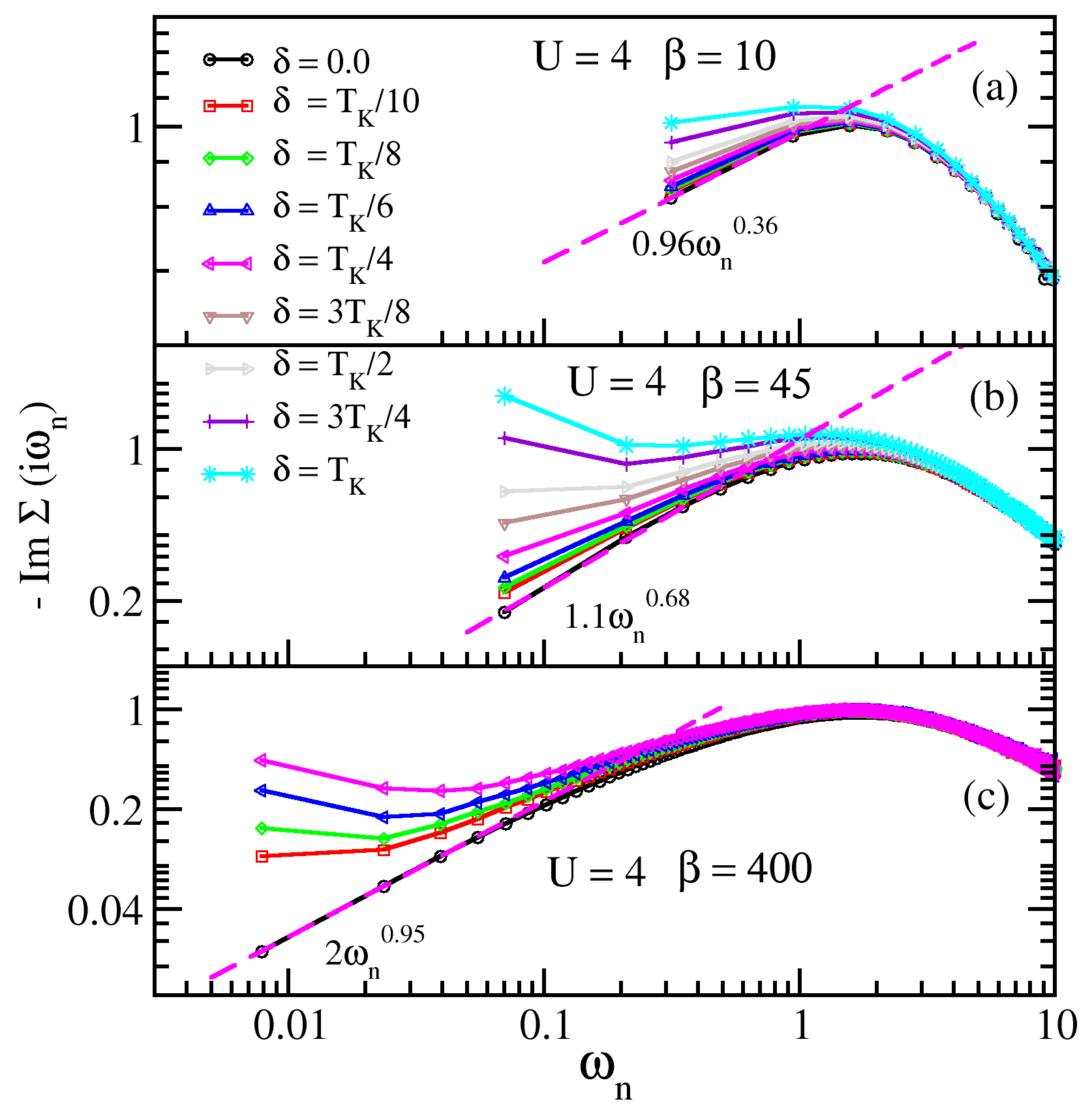}
\caption{(color online) Imaginary part of the Matsubara frequency self energy
for a range of values of the gap in the conduction bath density of states at
 (a) $\beta=10$, (b) $\beta=45$ and (c) $\beta=400$ with $U=4.0$ and $V=1.0$. 
 The dashed line is a power law fit to the low frequency limit of the gapless
 case. 
 %(d) Imaginary part of the self-energy for various $U$-values 
%for a fixed $\delta/T_K$=0.1 and $\beta$ = 500. Note the low frequency data
%collapse onto a universal scaling power law when plotted {\it vs} $\omega_n/T_K$ and scaled by a %multiplicative factor, $Y_f\sim {\cal{O}}(1)$.
}
\label{fig:fig1}
\end{figure}

\textit{Results and discussion.}$\textbf{---}$
The critical gap for the level crossing transition, from a singlet Fermi liquid
ground state to a doublet, is zero in the symmetric case~\cite{chen,logan2}. 
Hence at $T=0$, we expect a 
local moment ground state for any non-zero $\delta$. However, it is known
from $T=0$ LMA studies~\cite{logan2} that, although the low frequency single-particle spectrum of the gapped
case is very different from that of the $\delta=0$ case, the high frequency 
($\omega/T_K \gg \delta/T_K$) dynamics of the gapped system is identical
to that of the gapless case. Such a  crossover in the $T=0$ 
frequency dependence must manifest in a similar crossover in the temperature dependence.
Hence, for any finite gap, 
the system is expected to cross over from a local moment (LM) state to a generalized Fermi
liquid (GFL) with increasing temperature. We now explore
the finite temperature single- and two-particle quantities
in the GIAM to ascertain the existence and manifestation of such an LM to GFL crossover.

\begin{figure}[t]
\centering
\includegraphics[clip=,width=1.0\columnwidth]{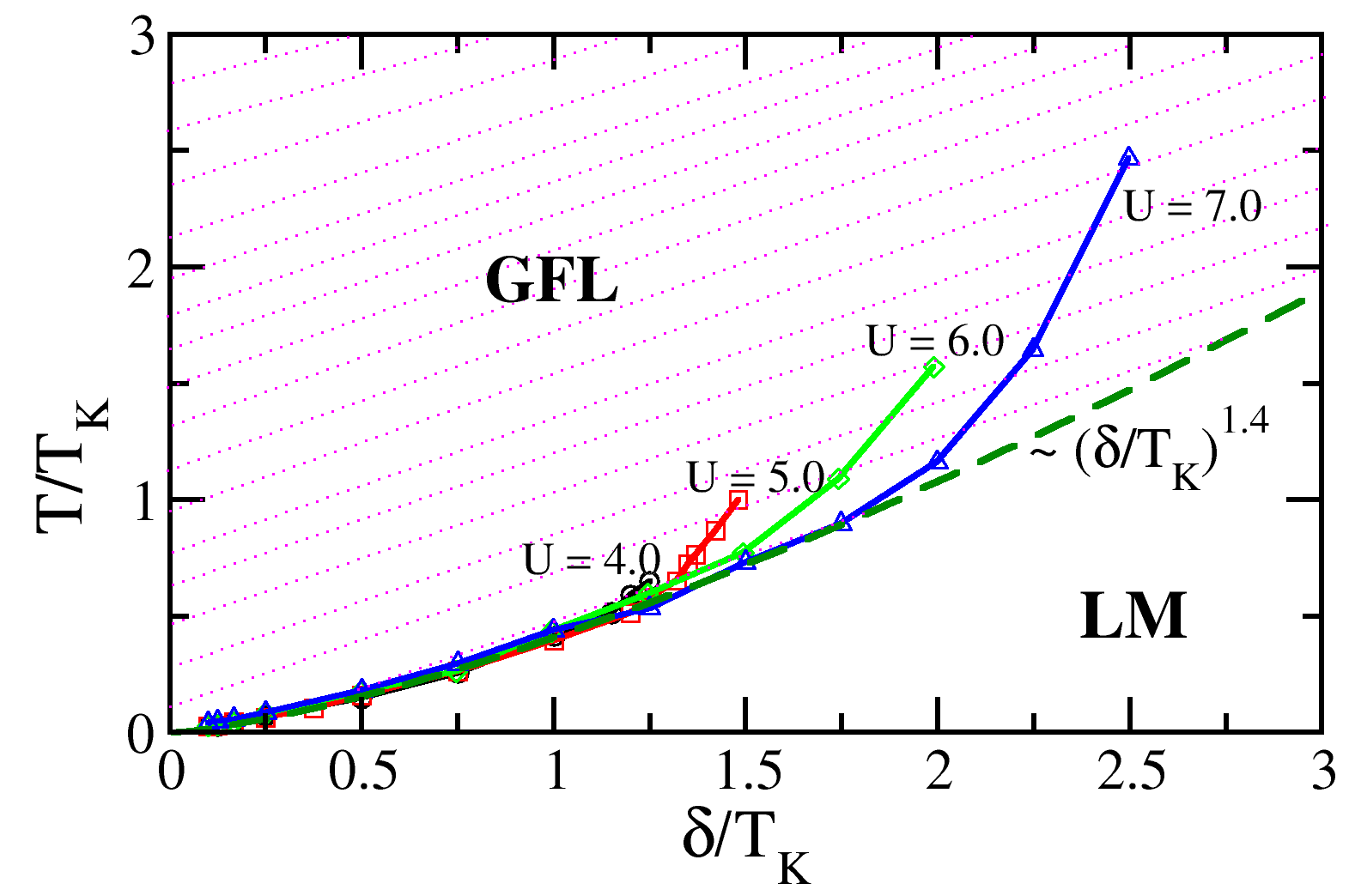}
\caption{(color online) A phase diagram in the 
$T/T_{K}$-$\delta/T_K$ plane for various $U$ values. The 
shaded region is the generalized Fermi liquid (GFL) while the unshaded one is the 
local moment (LM) region. The dashed line is the extrapolated, asymptotic strong 
coupling separatrix between the GFL and LM phases.}
\label{fig:fig2}
\end{figure}

The imaginary part of the self-energy is shown in Fig.~\ref{fig:fig1}  
for various gap values and decreasing temperature (from top to bottom)
for a fixed interaction strength, $U=4$. A low frequency power law is observed 
in the gapless case at all temperatures, the exponent of which approaches
unity as $T\rightarrow 0$. This is characteristic of  Fermi liquid 
formation in the  $\delta=0$ case. For the $\delta > 0 $ cases, although
at the lowest frequencies,  
the -${\rm Im}$ $\Sigma(i\omega_n)$ deviates from the  power law form of the gapless case,
it merges with the latter at higher $\omega_n$. 
%The bottom (d) panel of 
%Fig.~\ref{fig:fig1} shows $-{\rm Im}\Sigma(i\omega_n)$ {\it vs} $\omega_n/T_K$
%for various $U$-values, but a fixed $\delta/T_K$. The collapse onto a single power law
%with ${\cal{O}}(1)$ multiplicative factors indicates
%that, in the strong coupling limit, the exponent has a universal value, dependent
% only on $T/T_K$ and $\delta/T_K$. 
Furthermore,  Fig.~\ref{fig:fig1} shows that for lower gaps,
the deviation from the gapless case occurs at lower temperatures.
The temperature scale at which this change
in the $\omega_n$ dependence (from a power law form for $\delta=0$ to an upturn followed
by a power law for any $\delta>0$) occurs marks the crossover from a GFL to LM state and
is denoted by $T_{co}(\delta,U)$. We have determined the locus of such crossover temperatures
(see the SI\cite{SUPP} for details of the procedure) as a function of gap values for a given $U$
and  used it to construct a
`phase diagram' in the $T/T_K-\delta/T_K$  plane which is shown in Fig.~\ref{fig:fig2}. 

In  Fig.~\ref{fig:fig2} the region above the loci (for each $U$) represents the GFL, while the 
region below is the LM state. 
The universal, strong coupling asymptotic locus of the crossover points is the dashed
line in Fig.~\ref{fig:fig2}, which follows a form
$T_{co}=a(\delta/T_K)^b$ with $a\sim {\cal{O}}(1)$ and $b\sim 1.4$. 
In the limit of vanishing gap, the crossover temperature, $T_{co}\rightarrow 0$. This 
corroborates the result from earlier investigations~\cite{chen,logan2} 
that the critical gap for a local moment ground state is zero in the symmetric case.
 
\begin{figure}[t!]
\centering
\includegraphics[angle=0,width=1.0\columnwidth]{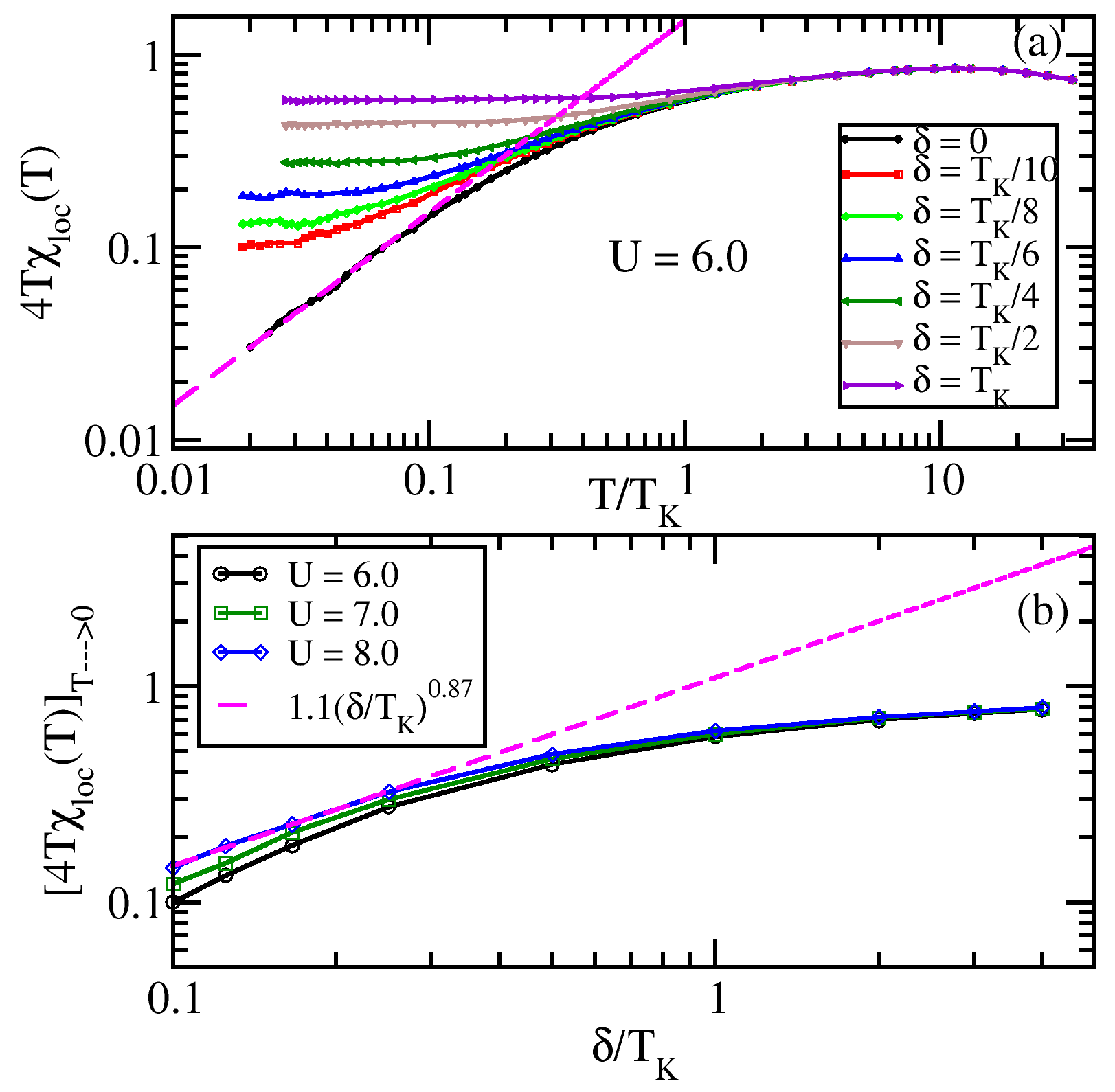}
\caption{(color online) (a) The product of temperature and 
the local static spin susceptibility (4T$\chi_{loc}(T)$) as a 
function of $T/T_K$ for a range of gap values at $U = 6.0$. 
The dashed line is a linear fit to the gapless case. 
(b) The $T\rightarrow 0$ residual moment on the impurity for different $U$ 
values as a function of gap. The brown dashed line is a power law fit to the small
gap region of the $U=8.0$ data.}
\label{fig:fig3}
\end{figure}

For the gapless case ($\delta$=0), the local static spin susceptibility, 
namely$;$ $\chi_{loc}(T)$=$\int^{\beta}_0 d\tau \langle S_z(\tau) S_z(0) \rangle$ 
is known\cite{HRK} to be temperature-independent for $T\ll T_K$, which represents 
Pauli-paramagnetic behaviour. Such behaviour indicates a complete screening of the local
moment. Nozieres proposed\cite{Nozieres} an exhaustion argument for heavy fermion 
systems, wherein one of the assumptions was that only those conduction electrons within 
an interval of $k_BT_K$ of the chemical potential are involved in the screening. However, 
it is now established\cite{slave-boson} that such an assumption is unjustified. The screening 
process involves electrons from infrared scales all the way to logarithmically high energy scales. 
Thus, with a gap in the vicinity of the chemical potential, we should expect that while the 
screening process {\em will} occur, the moment will not be completely screened. Indeed,
 this is seen in the upper panel of Fig.~\ref{fig:fig3} where we show $4T\chi_{loc}(T)$ for 
various gap fractions ($0.1 \leq \delta/T_K \leq 1$) as a function of temperature for a fixed $U=6.0$.
The gapless case (black symbols) shows a linear dependence (dashed line) as expected. 
However it must be noted that the linearity extends only up to about $T/T_K\sim 0.1$. For any finite gap,
it is seen that the low temperature $T\chi_{loc}(T)$ becomes flat indicating an unscreened moment, $m$ given by
$\lim_{T\rightarrow 0} (4T\chi_{loc}(T)) = m^2$. A higher gap would lead to a lesser number of conduction
 states  available for screening, hence the limiting zero temperature value of $m$ must increase with
 increasing $\delta$. This is shown in the lower panel of Fig.~\ref{fig:fig3}, where the square 
 of the moment {\it vs} $\delta/T_K$ is shown for three different $U$ values. A fit to the lower
 gap values indicates a power law dependence of $m^2$ on $\delta/T_K$ with the exponent $\sim 0.9$.
 We also note that, even with a large gap of $4T_K$, only about three-fourths of the moment is unscreened,
 hence states from non-universal scales are  involved in the Kondo screening of the magnetic moment.

\begin{figure}[t!]
\centering
\includegraphics[angle=0,width=1.0\columnwidth]{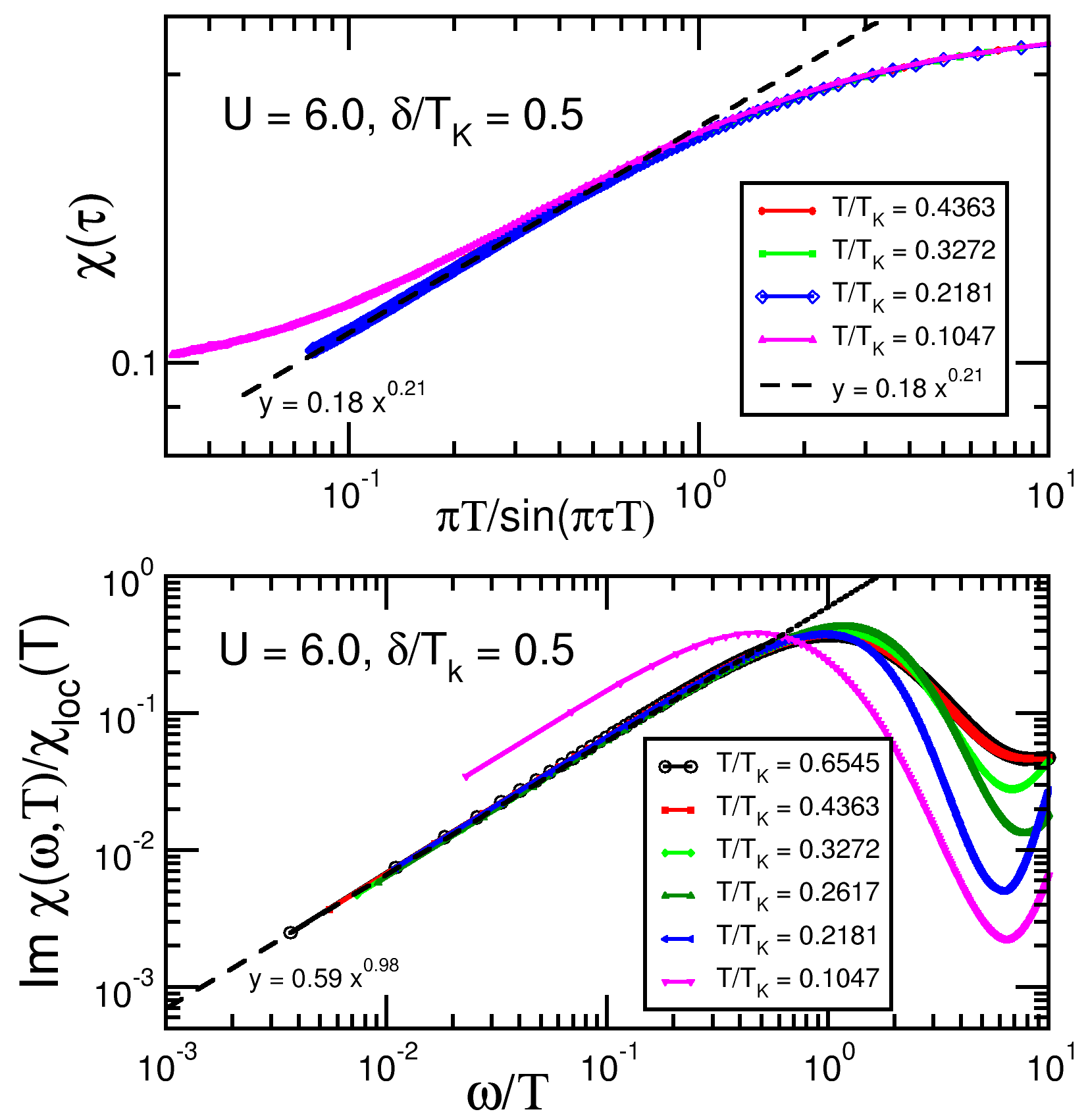}
\caption{(color online) Dynamical susceptibility scaling collapse: (a) $\chi(\tau)$ {\it vs}
$\pi T/\sin(\pi \tau T)$, (b) $\frac{{\rm Im}\chi(\omega)}{\chi_{loc}(T)}$ {\it vs.} $\omega/T$
for various temperature fractions with $U=6.0$ and $\delta/T_K=0.5$. The dashed lines in both the panels
 are power law fits.}
\label{fig:fig5}
\end{figure}

Since the critical gap for the quantum phase transition from a singlet to a 
doublet ground state\cite{Fradkin} is $\delta_c=0$, and the transition is continuous, 
we can expect a finite temperature critical scaling region, characterized by 
an $\omega/T$ scaling\cite{Ingersent} for real frequency quantities.
It has been shown through boundary conformal field 
theory arguments that such a scaling manifests as 
a $\pi T/\sin(\pi\tau T)$ scaling for imaginary time
quantities\cite{Ingersent,si,tsvelik}.
 In Fig.~\ref{fig:fig5},
we show the susceptibility $\chi(\tau)$ computed for $U=6$
and $\delta/T_K=1/2$ as a function of $\pi T/\sin(\pi \tau T)$ for various
 temperatures. A scaling collapse is evident for 
temperatures $T/T_K \gtrsim 0.218$, while for lower $T/T_K$,
a deviation from the power law scaling is observed. In the lower panel, 
a similar universal scaling collapse of the real frequency susceptibility
 (obtained through the maximum entropy method; see SI for details\cite{SUPP}) 
is observed when plotted as a function of $\omega/T$.
Such scaling behaviour has been observed previously in the pseudogap Anderson \cite{Ingersent} and Bose-Fermi Kondo models\cite{PhysRevLett.93.267201}. We note that the self-energy and static susceptibility show a crossover
from local moment like behaviour to generalized Fermi liquid behaviour at 
precisely the temperature above which the scaling collapse is observed
 (see Fig.~\ref{fig:fig1} and ~\ref{fig:fig2}). We have verified that the same holds for
other gaps as well (see figures 2 and 3 of SI\cite{SUPP}). 
Thus the shaded region of the finite temperature `phase diagram' shown in Fig.~\ref{fig:fig2}
is in fact the critical scaling region (or the `fan') of the
$\delta_c=0$ quantum critical point.  
 
\begin{figure}[t]
\centering
\includegraphics[angle=0,width=1.0\columnwidth]{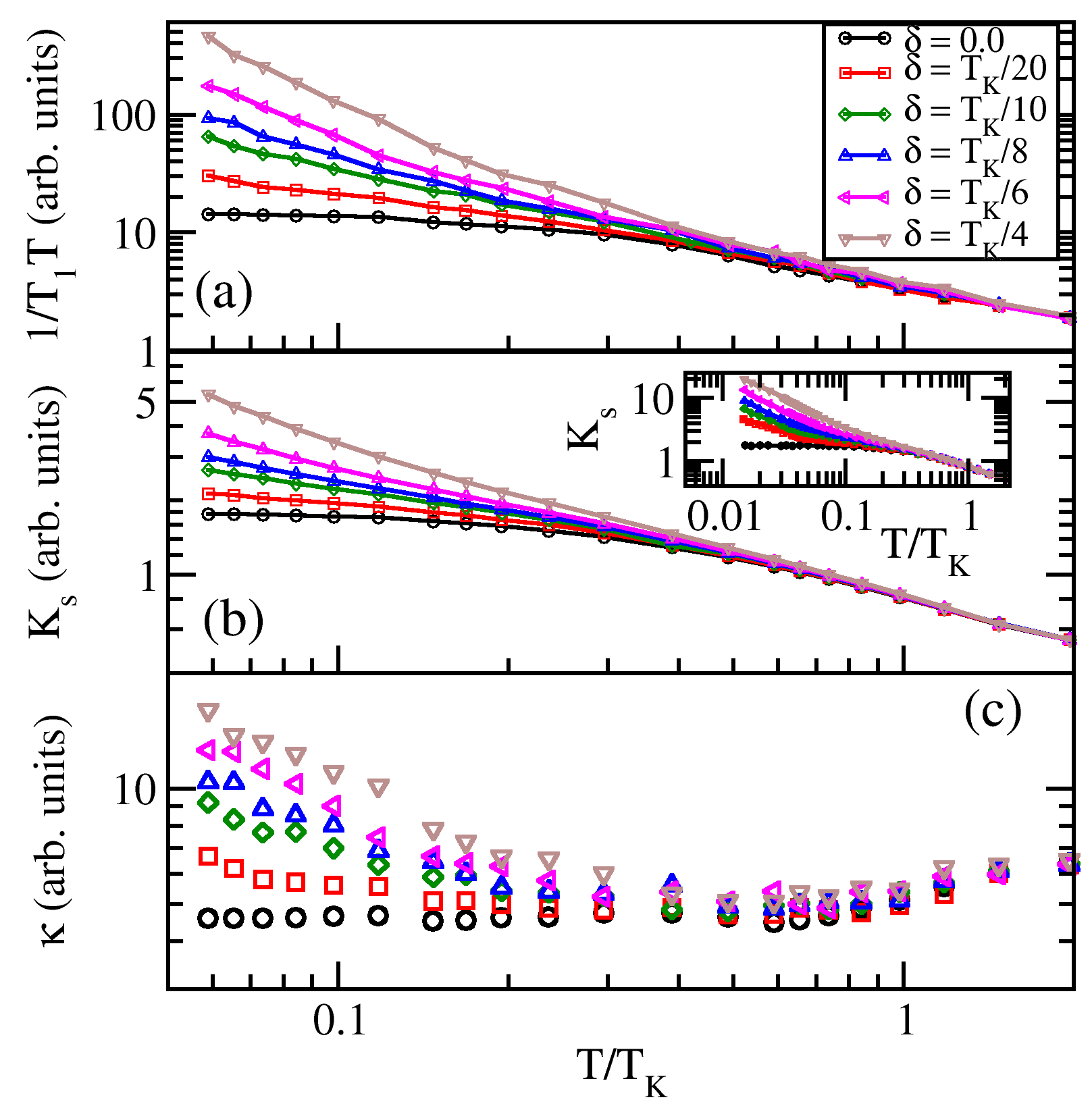}
\caption{(a) Magnetic relaxation rate, (b) Knight shift and 
(c) Korringa ratio as a function of temperature for various
 gap values and $U=4.0$.}
\label{fig:fig6}
\end{figure}

 The dynamical susceptibility $(\chi(\omega,T)=\chi'(\omega,T)+i\chi''(\omega,T))$ 
may be used directly to calculate experimentally measurable observables such as the 
 nuclear spin-lattice relaxation rate ($1/(T_1 T)$), Knight shift (K$_s$) and 
Korringa ratio ($\mathcal{K}$) (expressions provided in the SI\cite{SUPP}) 
\cite{Jarrell,PhysRevB.75.214515}.
%\begin{align}
%\frac{1}{T_1 T}  = A\lim_{\omega\rightarrow0}\frac{\chi''(\omega,T)}{\omega}
%\end{align}
%\vspace{-0.6cm}
%\begin{align}
%{\rm K}_s(T) = & B \chi'(0,T)= B \left[P \int 
%\frac{d\epsilon \chi''(\epsilon,T)}{\pi\epsilon}\right]\nonumber \\ & = B \chi_{loc}(T) \\
%\mathcal{K} &= \frac{C}{T_{1}T[K_{s}(T)]^{2}}\,.
%\end{align}
%Where A = $\frac{2k_{B}}{\gamma_{e}^{2}\hbar^{4}} |\tilde{A}|^2$, 
%B = $\frac{|\tilde{A}|}{\gamma_{e}\gamma_{n}\hbar^{2}}$ and 
%C = $\frac{\hbar}{4\pi k_{B}}(\frac{\gamma_{e}}{\gamma_{n}})^{2}$. 
%$\tilde{A}$ is the hyperfine coupling between the nuclear and electron spins, 
%and $\gamma_{n}$ ($\gamma_{e}$) is the nuclear (electronic) gyromagnetic ratio. 
%The main assumption in the above expressions is that the hyperfine coupling is momentum 
%independent. 
These three observables have been computed for various gap values
 and a fixed interaction strength ($U=4$) and are shown in fig.~\ref{fig:fig6} as a 
function of $T/T_K$. The singlet ground state in the gapless case implies that the 
relaxation mechanisms for the probe nuclear spin (e.g.\ $^{63}$Cu) due to the impurity
 spin (e.g.\ Fe) fluctuations would be suppressed sharply as
 the temperature drops below the Kondo scale. Thus the relaxation time scale should 
diverge with decreasing temperature. The result shown in the top panel of 
Fig.~\ref{fig:fig6} is in line with the expectations\cite{Pinto2012567}.
 For the gapless case,
 where the $1/T_1T$ saturates as $T\rightarrow 0$ 
implying that $T_1 \rightarrow \infty$. As seen from  Fig.~\ref{fig:fig3},
 the residual moment is finite for any non-zero gap, and moreover
 the magnitude of the moment increases with increasing 
 gap as $\sim (\delta/T_K)^{0.87}$. This would then imply that
 the coupling between the probe nuclear spin and the impurity 
 moment would remain finite even as $T\rightarrow 0$.
  For all $\delta \gtrsim T_K/10$, we find that the $1/(T_1T) \sim T^{-\alpha}$ with
 $\alpha > 1$ implying that $T_1 \sim T^{\alpha-1}$ and hence
 vanishes as $T\rightarrow 0$. However for $\delta=T_K/20$,
 we find that $\alpha \sim 0.67$, implying that $T_1$ diverges
 even though a residual moment exists. A diverging $T_1$ for
 a finite gap is surprising, and the origin of such a result
 is most likely that we need to go to even lower temperatures for smaller values of the gap to see the ground state behaviour. Nevertheless, the relaxation {\em rate} $1/(T_1T)$ 
 does diverge for any finite gap, and is hence consistent with
 the critical gap being zero in the symmetric case.
 
 The Knight shift is proportional to the static susceptibility, $\chi_{loc}(T)$.
 Hence, at temperatures below the Kondo scale in the gapless case, the $K_s$
 should saturate, which is indeed seen in the middle panel of Fig.~\ref{fig:fig6}.
 For any non-zero gap, the ground state being a doublet should yield a $1/T$ behaviour. 
For the higher gaps, the $1/T$ is clearly seen while for the lower gaps, much lower 
temperatures ($T\ll \delta$) need to be accessed to see such behaviour (see inset of the middle panel). Shiba has
considered the gapless Anderson impurity model\cite{Shiba01101975} and has proved
to all orders in perturbation theory that the Korringa ratio ($\kappa$) must be a 
constant as $T\rightarrow 0$. The bottom panel of Fig.~\ref{fig:fig6} confirms this, 
while showing that the $\kappa$
diverges with decreasing temperature for any finite gap in the host.

 In the present work, the manifestation of  
the zero gap quantum critical point in a precisely determined finite temperature region has been 
demonstrated through a striking scaling collapse of the dynamical susceptibility.
 We have also shown that this critical scaling region is characterized by
anomalous behaviour of various single-particle and two-particle static and dynamical quantities.
Based on dynamical spin susceptibility scaling as a function of $\omega/T$, we can classify the zero gap quantum
critical point as an interacting type QCP. The gapped Anderson impurity model is believed to be the
appropriate model for many material systems, such as dilute magnetic semiconductors and conventional superconductors. 
It could also be of potential relevance for lattice systems, where within the dynamical 
mean field theory framework, a gap could arise in the hybridization of the
self-consistently determined host. Our study yields 
an insight into the region and extent of the influence of
the zero gap quantum critical point on the finite temperature properties and hence could prove 
 to be important for the  understanding of such systems.

This work is supported by NSF DMR-1237565 and NSF EPSCoR Cooperative Agreement EPS-1003897 
with additional support from the Louisiana Board of Regents, and by CSIR and DST, India.
Our simulations used an open source 
implementation~\cite{Hafer} of the hybridization expansion continuous-time quantum Monte 
Carlo algorithm~\cite{CTQMC1} and the ALPS~\cite{Bauer} libraries. 
Supercomputer support is 
provided by the Louisiana Optical Network Initiative (LONI) and HPC@LSU.
We acknowledge Sandeep Kumar Reddy for his support on the installation of ALPS.
SA acknowledges JNCASR for support during his visits.

\bibliographystyle{apsrev4-1}
\bibliography{apssamp}

%Merlin.mbs v4.21 2009-07-09.
\begin{thebibliography}{10}%
\makeatletter
\providecommand \@ifxundefined [1]{%
 \ifx #1\undefined \expandafter \@firstoftwo
 \else \expandafter \@secondoftwo
\fi
}%
\providecommand \@ifnum [1]{%
 \ifnum #1\expandafter \@firstoftwo
 \else \expandafter \@secondoftwo
\fi
}%
\providecommand \enquote [1]{``#1''}%
\providecommand \bibnamefont  [1]{#1}%
\providecommand \bibfnamefont [1]{#1}%
\providecommand \citenamefont [1]{#1}%
\providecommand\href[0]{\@sanitize\@href}%
\providecommand\@href[1]{\endgroup\@@startlink{#1}\endgroup\@@href}%
\providecommand\@@href[1]{#1\@@endlink}%
\providecommand \@sanitize [0]{\begingroup\catcode`\&12\catcode`\#12\relax}%
\@ifxundefined \pdfoutput {\@firstoftwo}{%
 \@ifnum{\z@=\pdfoutput}{\@firstoftwo}{\@secondoftwo}%
}{%
 \providecommand\@@startlink[1]{\leavevmode\special{html:<a href="#1">}}%
 \providecommand\@@endlink[0]{\special{html:</a>}}%
}{%
 \providecommand\@@startlink[1]{%
  \leavevmode
  \pdfstartlink
   attr{/Border[0 0 1 ]/H/I/C[0 1 1]}%
   user{/Subtype/Link/A<</Type/Action/S/URI/URI(#1)>>}%
  \relax
 }%
 \providecommand\@@endlink[0]{\pdfendlink}%
}%
\providecommand \url  [0]{\begingroup\@sanitize \@url }%
\providecommand \@url [1]{\endgroup\@href {#1}{\urlprefix}}%
\providecommand \urlprefix [0]{URL }%
\providecommand \Eprint[0]{\href }%
\@ifxundefined \urlstyle {%
  \providecommand \doi [1]{doi:\discretionary{}{}{}#1}%
}{%
  \providecommand \doi [0]{doi:\discretionary{}{}{}\begingroup
  \urlstyle{rm}\Url }%
}%
\providecommand \doibase [0]{http://dx.doi.org/}%
\providecommand \Doi[1]{\href{\doibase#1}}%
\providecommand \bibAnnote [3]{%
  \BibitemShut{#1}%
  \begin{quotation}\noindent
    \textsc{Key:}\ #2\\\textsc{Annotation:}\ #3%
  \end{quotation}%
}%
\providecommand \bibAnnoteFile [2]{%
  \IfFileExists{#2}{\bibAnnote {#1} {#2} {\input{#2}}}{}%
}%
\providecommand \typeout [0]{\immediate \write \m@ne }%
\providecommand \selectlanguage [0]{\@gobble}%
\providecommand \bibinfo [0]{\@secondoftwo}%
\providecommand \bibfield [0]{\@secondoftwo}%
\providecommand \translation [1]{[#1]}%
\providecommand \BibitemOpen[0]{}%
\providecommand \bibitemStop [0]{}%
\providecommand \bibitemNoStop [0]{.\EOS\space}%
\providecommand \EOS [0]{\spacefactor3000\relax}%
\providecommand \BibitemShut [1]{\csname bibitem#1\endcsname}%
%</preamble>
\bibitem{kondo}%
  \BibitemOpen
  \bibfield{author}{%
  \bibinfo {author} {\bibfnamefont{J.}~\bibnamefont{Kondo}},\ }%
  \bibfield{journal}{%
  \Doi{10.1143/PTP.32.37}{\bibinfo {journal} {Progress of Theoretical
  Physics}}\ }%
  \textbf{\bibinfo {volume} {32}},\ \bibinfo {pages} {37} (\bibinfo {year}
  {1964})%
  \bibAnnoteFile{NoStop}{kondo}%
\bibitem{dot}%
  \BibitemOpen
  \bibfield{author}{%
  \bibinfo {author} {\bibfnamefont{D.}~\bibnamefont{Goldhaber-Gordon}},
  \bibinfo {author} {\bibfnamefont{H.}~\bibnamefont{Shtrikman}}, \bibinfo
  {author} {\bibfnamefont{D.}~\bibnamefont{Mahalu}}, \bibinfo {author}
  {\bibfnamefont{D.}~\bibnamefont{Abusch-Magder}}, \bibinfo {author}
  {\bibfnamefont{U.}~\bibnamefont{Meirav}},\ and\ \bibinfo {author}
  {\bibfnamefont{M.}~\bibnamefont{Kastner}},\ }%
  \bibfield{journal}{%
  \bibinfo {journal} {Nature}\ }%
  \textbf{\bibinfo {volume} {391}},\ \bibinfo {pages} {156} (\bibinfo {year}
  {1998})%
  \bibAnnoteFile{NoStop}{dot}%
\bibitem{PhysRevLett.89.256801}%
  \BibitemOpen
  \bibfield{author}{%
  \bibinfo {author} {\bibfnamefont{M.~R.}\ \bibnamefont{Buitelaar}}, \bibinfo
  {author} {\bibfnamefont{T.}~\bibnamefont{Nussbaumer}},\ and\ \bibinfo
  {author} {\bibfnamefont{C.}~\bibnamefont{Sch\"onenberger}},\ }%
  \bibfield{journal}{%
  \Doi{10.1103/PhysRevLett.89.256801}{\bibinfo {journal} {Phys. Rev. Lett.}}\
  }%
  \textbf{\bibinfo {volume} {89}},\ \bibinfo {pages} {256801} (\bibinfo {month}
  {Dec}\ \bibinfo {year} {2002}),\
  \url{http://link.aps.org/doi/10.1103/PhysRevLett.89.256801}%
  \bibAnnoteFile{NoStop}{PhysRevLett.89.256801}%
\bibitem{PhysRevB.88.045101}%
  \BibitemOpen
  \bibfield{author}{%
  \bibinfo {author} {\bibfnamefont{J.-D.}\ \bibnamefont{Pillet}}, \bibinfo
  {author} {\bibfnamefont{P.}~\bibnamefont{Joyez}}, \bibinfo {author}
  {\bibfnamefont{R.}~\bibnamefont{\ifmmode~\check{Z}\else \v{Z}\fi{}itko}},\
  and\ \bibinfo {author} {\bibfnamefont{M.~F.}\ \bibnamefont{Goffman}},\ }%
  \bibfield{journal}{%
  \Doi{10.1103/PhysRevB.88.045101}{\bibinfo {journal} {Phys. Rev. B}}\ }%
  \textbf{\bibinfo {volume} {88}},\ \bibinfo {pages} {045101} (\bibinfo {month}
  {Jul}\ \bibinfo {year} {2013}),\
  \url{http://link.aps.org/doi/10.1103/PhysRevB.88.045101}%
  \bibAnnoteFile{NoStop}{PhysRevB.88.045101}%
\bibitem{PhysRevB.92.161108}%
  \BibitemOpen
  \bibfield{author}{%
  \bibinfo {author} {\bibfnamefont{M.}~\bibnamefont{Okawa}}, \bibinfo {author}
  {\bibfnamefont{Y.}~\bibnamefont{Ishida}}, \bibinfo {author}
  {\bibfnamefont{M.}~\bibnamefont{Takahashi}}, \bibinfo {author}
  {\bibfnamefont{T.}~\bibnamefont{Shimada}}, \bibinfo {author}
  {\bibfnamefont{F.}~\bibnamefont{Iga}}, \bibinfo {author}
  {\bibfnamefont{T.}~\bibnamefont{Takabatake}}, \bibinfo {author}
  {\bibfnamefont{T.}~\bibnamefont{Saitoh}},\ and\ \bibinfo {author}
  {\bibfnamefont{S.}~\bibnamefont{Shin}},\ }%
  \bibfield{journal}{%
  \Doi{10.1103/PhysRevB.92.161108}{\bibinfo {journal} {Phys. Rev. B}}\ }%
  \textbf{\bibinfo {volume} {92}},\ \bibinfo {pages} {161108} (\bibinfo {month}
  {Oct}\ \bibinfo {year} {2015}),\
  \url{http://link.aps.org/doi/10.1103/PhysRevB.92.161108}%
  \bibAnnoteFile{NoStop}{PhysRevB.92.161108}%
\bibitem{PhysRevLett.112.136401}%
  \BibitemOpen
  \bibfield{author}{%
  \bibinfo {author} {\bibfnamefont{W.}~\bibnamefont{Ruan}}, \bibinfo {author}
  {\bibfnamefont{C.}~\bibnamefont{Ye}}, \bibinfo {author}
  {\bibfnamefont{M.}~\bibnamefont{Guo}}, \bibinfo {author}
  {\bibfnamefont{F.}~\bibnamefont{Chen}}, \bibinfo {author}
  {\bibfnamefont{X.}~\bibnamefont{Chen}}, \bibinfo {author}
  {\bibfnamefont{G.-M.}\ \bibnamefont{Zhang}},\ and\ \bibinfo {author}
  {\bibfnamefont{Y.}~\bibnamefont{Wang}},\ }%
  \bibfield{journal}{%
  \Doi{10.1103/PhysRevLett.112.136401}{\bibinfo {journal} {Phys. Rev. Lett.}}\
  }%
  \textbf{\bibinfo {volume} {112}},\ \bibinfo {pages} {136401} (\bibinfo
  {month} {Mar}\ \bibinfo {year} {2014}),\
  \url{http://link.aps.org/doi/10.1103/PhysRevLett.112.136401}%
  \bibAnnoteFile{NoStop}{PhysRevLett.112.136401}%
\bibitem{2108131}%
  \BibitemOpen
  \bibfield{author}{%
  \bibinfo {author} {\bibfnamefont{H.~T.}\ \bibnamefont{He}}, \bibinfo {author}
  {\bibfnamefont{C.~L.}\ \bibnamefont{Yang}}, \bibinfo {author}
  {\bibfnamefont{W.~K.}\ \bibnamefont{Ge}}, \bibinfo {author}
  {\bibfnamefont{J.~N.}\ \bibnamefont{Wang}}, \bibinfo {author}
  {\bibfnamefont{X.}~\bibnamefont{Dai}},\ and\ \bibinfo {author}
  {\bibfnamefont{Y.~Q.}\ \bibnamefont{Wang}},\ }%
  \bibfield{journal}{%
  \bibinfo {journal} {Applied Physics Letters}\ }%
  \textbf{\bibinfo {volume} {87}},\ \bibinfo {eid} {162506} (\bibinfo {year}
  {2005}),\
  \url{http://scitation.aip.org/content/aip/journal/apl/87/16/10.1063/1.210813%
1}%
  \bibAnnoteFile{NoStop}{2108131}%
\bibitem{PhysRevB.76.195207}%
  \BibitemOpen
  \bibfield{author}{%
  \bibinfo {author} {\bibfnamefont{R.}~\bibnamefont{Ramaneti}}, \bibinfo
  {author} {\bibfnamefont{J.~C.}\ \bibnamefont{Lodder}},\ and\ \bibinfo
  {author} {\bibfnamefont{R.}~\bibnamefont{Jansen}},\ }%
  \bibfield{journal}{%
  \Doi{10.1103/PhysRevB.76.195207}{\bibinfo {journal} {Phys. Rev. B}}\ }%
  \textbf{\bibinfo {volume} {76}},\ \bibinfo {pages} {195207} (\bibinfo {month}
  {Nov}\ \bibinfo {year} {2007}),\
  \url{http://link.aps.org/doi/10.1103/PhysRevB.76.195207}%
  \bibAnnoteFile{NoStop}{PhysRevB.76.195207}%
\bibitem{saso1}%
  \BibitemOpen
  \bibfield{author}{%
  \bibinfo {author} {\bibfnamefont{T.}~\bibnamefont{Saso}},\ }%
  \bibfield{journal}{%
  \bibinfo {journal} {Journal of the Physical Society of Japan}\ }%
  \textbf{\bibinfo {volume} {61}},\ \bibinfo {pages} {3439} (\bibinfo {year}
  {1992}),\ \url{http://journals.jps.jp/doi/abs/10.1143/JPSJ.61.3439}%
  \bibAnnoteFile{NoStop}{saso1}%
\bibitem{Takegahara}%
  \BibitemOpen
  \bibfield{author}{%
  \bibinfo {author} {\bibfnamefont{K.}~\bibnamefont{Takegahara}}, \bibinfo
  {author} {\bibfnamefont{Y.}~\bibnamefont{Shimizu}},\ and\ \bibinfo {author}
  {\bibfnamefont{O.}~\bibnamefont{Sakai}},\ }%
  \bibfield{journal}{%
  \Doi{10.1143/JPSJ.61.3443}{\bibinfo {journal} {Journal of the Physical
  Society of Japan}}\ }%
  \textbf{\bibinfo {volume} {61}},\ \bibinfo {pages} {3443} (\bibinfo {year}
  {1992}),\ \url{http://dx.doi.org/10.1143/JPSJ.61.3443}%
  \bibAnnoteFile{NoStop}{Takegahara}%
\bibitem{saso2}%
  \BibitemOpen
  \bibfield{author}{%
  \bibinfo {author} {\bibfnamefont{J.}~\bibnamefont{Ogura}}\ and\ \bibinfo
  {author} {\bibfnamefont{T.}~\bibnamefont{Saso}},\ }%
  \bibfield{journal}{%
  \Doi{10.1143/JPSJ.62.4364}{\bibinfo {journal} {Journal of the Physical
  Society of Japan}}\ }%
  \textbf{\bibinfo {volume} {62}},\ \bibinfo {pages} {4364} (\bibinfo {year}
  {1993})%
  \bibAnnoteFile{NoStop}{saso2}%
\bibitem{chen}%
  \BibitemOpen
  \bibfield{author}{%
  \bibinfo {author} {\bibfnamefont{K.}~\bibnamefont{Chen}}\ and\ \bibinfo
  {author} {\bibfnamefont{C.}~\bibnamefont{Jayaprakash}},\ }%
  \bibfield{journal}{%
  \Doi{10.1103/PhysRevB.57.5225}{\bibinfo {journal} {Phys. Rev. B}}\ }%
  \textbf{\bibinfo {volume} {57}},\ \bibinfo {pages} {5225} (\bibinfo {month}
  {Mar}\ \bibinfo {year} {1998}),\
  \url{http://link.aps.org/doi/10.1103/PhysRevB.57.5225}%
  \bibAnnoteFile{NoStop}{chen}%
\bibitem{moca}%
  \BibitemOpen
  \bibfield{author}{%
  \bibinfo {author} {\bibfnamefont{C.~P.}\ \bibnamefont{Moca}}\ and\ \bibinfo
  {author} {\bibfnamefont{A.}~\bibnamefont{Roman}},\ }%
  \bibfield{journal}{%
  \Doi{10.1103/PhysRevB.81.235106}{\bibinfo {journal} {Phys. Rev. B}}\ }%
  \textbf{\bibinfo {volume} {81}},\ \bibinfo {pages} {235106} (\bibinfo {month}
  {Jun}\ \bibinfo {year} {2010}),\
  \url{http://link.aps.org/doi/10.1103/PhysRevB.81.235106}%
  \bibAnnoteFile{NoStop}{moca}%
\bibitem{Pinto2012567}%
  \BibitemOpen
  \bibfield{author}{%
  \bibinfo {author} {\bibfnamefont{J.}~\bibnamefont{Pinto}}\ and\ \bibinfo
  {author} {\bibfnamefont{H.}~\bibnamefont{Frota}},\ }%
  \bibfield{journal}{%
  \bibinfo {journal} {Solid State Communications}\ }%
  \textbf{\bibinfo {volume} {152}},\ \bibinfo {pages} {567 } (\bibinfo {year}
  {2012}),\ ISSN \bibinfo {issn} {0038-1098},\
  \url{http://www.sciencedirect.com/science/article/pii/S0038109812000543}%
  \bibAnnoteFile{NoStop}{Pinto2012567}%
\bibitem{logan2}%
  \BibitemOpen
  \bibfield{author}{%
  \bibinfo {author} {\bibfnamefont{M.~R.}\ \bibnamefont{Galpin}}\ and\ \bibinfo
  {author} {\bibfnamefont{D.~E.}\ \bibnamefont{Logan}},\ }%
  \bibfield{journal}{%
  \Doi{10.1140/epjb/e2008-00138-5}{\bibinfo {journal} {Eur. Phys. J. B}}\ }%
  \textbf{\bibinfo {volume} {62}},\ \bibinfo {pages} {129} (\bibinfo {year}
  {2008}),\ \url{http://dx.doi.org/10.1140/epjb/e2008-00138-5}%
  \bibAnnoteFile{NoStop}{logan2}%
\bibitem{logan1}%
  \BibitemOpen
  \bibfield{author}{%
  \bibinfo {author} {\bibfnamefont{M.~R.}\ \bibnamefont{Galpin}}\ and\ \bibinfo
  {author} {\bibfnamefont{D.~E.}\ \bibnamefont{Logan}},\ }%
  \bibfield{journal}{%
  \Doi{10.1103/PhysRevB.77.195108}{\bibinfo {journal} {Phys. Rev. B}}\ }%
  \textbf{\bibinfo {volume} {77}},\ \bibinfo {pages} {195108} (\bibinfo {month}
  {May}\ \bibinfo {year} {2008}),\
  \url{http://link.aps.org/doi/10.1103/PhysRevB.77.195108}%
  \bibAnnoteFile{NoStop}{logan1}%
\bibitem{Fradkin}%
  \BibitemOpen
  \bibfield{author}{%
  \bibinfo {author} {\bibfnamefont{D.}~\bibnamefont{Withoff}}\ and\ \bibinfo
  {author} {\bibfnamefont{E.}~\bibnamefont{Fradkin}},\ }%
  \bibfield{journal}{%
  \Doi{10.1103/PhysRevLett.64.1835}{\bibinfo {journal} {Phys. Rev. Lett.}}\ }%
  \textbf{\bibinfo {volume} {64}},\ \bibinfo {pages} {1835} (\bibinfo {month}
  {Apr}\ \bibinfo {year} {1990}),\
  \url{http://link.aps.org/doi/10.1103/PhysRevLett.64.1835}%
  \bibAnnoteFile{NoStop}{Fradkin}%
\bibitem{Ingersent}%
  \BibitemOpen
  \bibfield{author}{%
  \bibinfo {author} {\bibfnamefont{K.}~\bibnamefont{Ingersent}}\ and\ \bibinfo
  {author} {\bibfnamefont{Q.}~\bibnamefont{Si}},\ }%
  \bibfield{journal}{%
  \Doi{10.1103/PhysRevLett.89.076403}{\bibinfo {journal} {Phys. Rev. Lett.}}\
  }%
  \textbf{\bibinfo {volume} {89}},\ \bibinfo {pages} {076403} (\bibinfo {month}
  {Jul}\ \bibinfo {year} {2002}),\
  \url{http://link.aps.org/doi/10.1103/PhysRevLett.89.076403}%
  \bibAnnoteFile{NoStop}{Ingersent}%
\bibitem{si}%
  \BibitemOpen
  \bibfield{author}{%
  \bibinfo {author} {\bibfnamefont{S.}~\bibnamefont{Kirchner}}\ and\ \bibinfo
  {author} {\bibfnamefont{Q.}~\bibnamefont{Si}},\ }%
  \bibfield{journal}{%
  \Doi{10.1103/PhysRevLett.100.026403}{\bibinfo {journal} {Phys. Rev. Lett.}}\
  }%
  \textbf{\bibinfo {volume} {100}},\ \bibinfo {pages} {026403} (\bibinfo
  {month} {Jan}\ \bibinfo {year} {2008}),\
  \url{http://link.aps.org/doi/10.1103/PhysRevLett.100.026403}%
  \bibAnnoteFile{NoStop}{si}%
\bibitem{tsvelik}%
  \BibitemOpen
  \bibfield{author}{%
  \bibinfo {author} {\bibfnamefont{A.}~\bibnamefont{Tsvelik}},\ }%
  \enquote{\bibinfo {title} {Quantum field theory in condensed matter
  physics},}\ \ (\bibinfo {publisher} {Cambridge University Press, Cambridge,
  England},\ \bibinfo {year} {1996})%
  \bibAnnoteFile{NoStop}{tsvelik}%
\bibitem{10.1080}%
  \BibitemOpen
  \bibfield{author}{%
  \bibinfo {author} {\bibfnamefont{M.}~\bibnamefont{Vojta}},\ }%
  \bibfield{journal}{%
  \Doi{10.1080/14786430500070396}{\bibinfo {journal} {Philosophical Magazine}}\
  }%
  \textbf{\bibinfo {volume} {86}},\ \bibinfo {pages} {1807} (\bibinfo {year}
  {2006}),\
  \Eprint{http://arxiv.org/abs/http://dx.doi.org/10.1080/14786430500070396}{ht%
tp://dx.doi.org/10.1080/14786430500070396},\
  \url{http://dx.doi.org/10.1080/14786430500070396}%
  \bibAnnoteFile{NoStop}{10.1080}%
\bibitem{PhysRevLett.107.076404}%
  \BibitemOpen
  \bibfield{author}{%
  \bibinfo {author} {\bibfnamefont{M.~T.}\ \bibnamefont{Glossop}}, \bibinfo
  {author} {\bibfnamefont{S.}~\bibnamefont{Kirchner}}, \bibinfo {author}
  {\bibfnamefont{J.~H.}\ \bibnamefont{Pixley}},\ and\ \bibinfo {author}
  {\bibfnamefont{Q.}~\bibnamefont{Si}},\ }%
  \bibfield{journal}{%
  \Doi{10.1103/PhysRevLett.107.076404}{\bibinfo {journal} {Phys. Rev. Lett.}}\
  }%
  \textbf{\bibinfo {volume} {107}},\ \bibinfo {pages} {076404} (\bibinfo
  {month} {Aug}\ \bibinfo {year} {2011}),\
  \url{http://link.aps.org/doi/10.1103/PhysRevLett.107.076404}%
  \bibAnnoteFile{NoStop}{PhysRevLett.107.076404}%
\bibitem{PhysRevB.91.035118}%
  \BibitemOpen
  \bibfield{author}{%
  \bibinfo {author} {\bibfnamefont{T.}~\bibnamefont{Chowdhury}}\ and\ \bibinfo
  {author} {\bibfnamefont{K.}~\bibnamefont{Ingersent}},\ }%
  \bibfield{journal}{%
  \Doi{10.1103/PhysRevB.91.035118}{\bibinfo {journal} {Phys. Rev. B}}\ }%
  \textbf{\bibinfo {volume} {91}},\ \bibinfo {pages} {035118} (\bibinfo {month}
  {Jan}\ \bibinfo {year} {2015}),\
  \url{http://link.aps.org/doi/10.1103/PhysRevB.91.035118}%
  \bibAnnoteFile{NoStop}{PhysRevB.91.035118}%
\bibitem{gegenwart2008quantum}%
  \BibitemOpen
  \bibfield{author}{%
  \bibinfo {author} {\bibfnamefont{P.}~\bibnamefont{Gegenwart}}, \bibinfo
  {author} {\bibfnamefont{Q.}~\bibnamefont{Si}},\ and\ \bibinfo {author}
  {\bibfnamefont{F.}~\bibnamefont{Steglich}},\ }%
  \bibfield{journal}{%
  \bibinfo {journal} {nature physics}\ }%
  \textbf{\bibinfo {volume} {4}},\ \bibinfo {pages} {186} (\bibinfo {year}
  {2008})%
  \bibAnnoteFile{NoStop}{gegenwart2008quantum}%
\bibitem{CTQMC1}%
  \BibitemOpen
  \bibfield{author}{%
  \bibinfo {author} {\bibfnamefont{P.}~\bibnamefont{Werner}}, \bibinfo {author}
  {\bibfnamefont{A.}~\bibnamefont{Comanac}}, \bibinfo {author}
  {\bibfnamefont{L.}~\bibnamefont{de' Medici}}, \bibinfo {author}
  {\bibfnamefont{M.}~\bibnamefont{Troyer}},\ and\ \bibinfo {author}
  {\bibfnamefont{A.~J.}\ \bibnamefont{Millis}},\ }%
  \bibfield{journal}{%
  \Doi{10.1103/PhysRevLett.97.076405}{\bibinfo {journal} {Phys. Rev. Lett.}}\
  }%
  \textbf{\bibinfo {volume} {97}},\ \bibinfo {pages} {076405} (\bibinfo {month}
  {Aug}\ \bibinfo {year} {2006}),\
  \url{http://link.aps.org/doi/10.1103/PhysRevLett.97.076405}%
  \bibAnnoteFile{NoStop}{CTQMC1}%
\bibitem{Jarrell}%
  \BibitemOpen
  \bibfield{author}{%
  \bibinfo {author} {\bibfnamefont{M.}~\bibnamefont{Jarrell}}, \bibinfo
  {author} {\bibfnamefont{J.~E.}\ \bibnamefont{Gubernatis}},\ and\ \bibinfo
  {author} {\bibfnamefont{R.~N.}\ \bibnamefont{Silver}},\ }%
  \bibfield{journal}{%
  \Doi{10.1103/PhysRevB.44.5347}{\bibinfo {journal} {Phys. Rev. B}}\ }%
  \textbf{\bibinfo {volume} {44}},\ \bibinfo {pages} {5347} (\bibinfo {month}
  {Sep}\ \bibinfo {year} {1991}),\
  \url{http://link.aps.org/doi/10.1103/PhysRevB.44.5347}%
  \bibAnnoteFile{NoStop}{Jarrell}%
\bibitem{SUPP}%
  \BibitemOpen
  \bibinfo {note} {See Supplemental Material at [URL will be inserted by
  publisher] for more details about the procedure for extraction of finite
  temperature phase diagram, maximum entropy method and scaling curves of
  dynamical susceptibilty in the critical region}%
  \bibAnnoteFile{NoStop}{SUPP}%
\bibitem{HRK}%
  \BibitemOpen
  \bibfield{author}{%
  \bibinfo {author} {\bibfnamefont{H.~R.}\ \bibnamefont{Krishna-murthy}},
  \bibinfo {author} {\bibfnamefont{J.~W.}\ \bibnamefont{Wilkins}},\ and\
  \bibinfo {author} {\bibfnamefont{K.~G.}\ \bibnamefont{Wilson}},\ }%
  \bibfield{journal}{%
  \Doi{10.1103/PhysRevB.21.1003}{\bibinfo {journal} {Phys. Rev. B}}\ }%
  \textbf{\bibinfo {volume} {21}},\ \bibinfo {pages} {1003} (\bibinfo {month}
  {Feb}\ \bibinfo {year} {1980}),\
  \url{http://link.aps.org/doi/10.1103/PhysRevB.21.1003}%
  \bibAnnoteFile{NoStop}{HRK}%
\bibitem{Nozieres}%
  \BibitemOpen
  \bibfield{author}{%
  \bibinfo {author} {\bibfnamefont{P.}~\bibnamefont{Nozières}},\ }%
  \bibfield{journal}{%
  \Doi{10.1007/s100510050571}{\bibinfo {journal} {The European Physical Journal
  B - Condensed Matter and Complex Systems}}\ }%
  \textbf{\bibinfo {volume} {6}},\ \bibinfo {pages} {447} (\bibinfo {year}
  {1998}),\ ISSN \bibinfo {issn} {1434-6028},\
  \url{http://dx.doi.org/10.1007/s100510050571}%
  \bibAnnoteFile{NoStop}{Nozieres}%
\bibitem{slave-boson}%
  \BibitemOpen
  \bibfield{author}{%
  \bibinfo {author} {\bibfnamefont{S.}~\bibnamefont{Burdin}}, \bibinfo {author}
  {\bibfnamefont{A.}~\bibnamefont{Georges}},\ and\ \bibinfo {author}
  {\bibfnamefont{D.~R.}\ \bibnamefont{Grempel}},\ }%
  \bibfield{journal}{%
  \Doi{10.1103/PhysRevLett.85.1048}{\bibinfo {journal} {Phys. Rev. Lett.}}\ }%
  \textbf{\bibinfo {volume} {85}},\ \bibinfo {pages} {1048} (\bibinfo {month}
  {Jul}\ \bibinfo {year} {2000}),\
  \url{http://link.aps.org/doi/10.1103/PhysRevLett.85.1048}%
  \bibAnnoteFile{NoStop}{slave-boson}%
\bibitem{PhysRevLett.93.267201}%
  \BibitemOpen
  \bibfield{author}{%
  \bibinfo {author} {\bibfnamefont{L.}~\bibnamefont{Zhu}}, \bibinfo {author}
  {\bibfnamefont{S.}~\bibnamefont{Kirchner}}, \bibinfo {author}
  {\bibfnamefont{Q.}~\bibnamefont{Si}},\ and\ \bibinfo {author}
  {\bibfnamefont{A.}~\bibnamefont{Georges}},\ }%
  \bibfield{journal}{%
  \Doi{10.1103/PhysRevLett.93.267201}{\bibinfo {journal} {Phys. Rev. Lett.}}\
  }%
  \textbf{\bibinfo {volume} {93}},\ \bibinfo {pages} {267201} (\bibinfo {month}
  {Dec}\ \bibinfo {year} {2004}),\
  \url{http://link.aps.org/doi/10.1103/PhysRevLett.93.267201}%
  \bibAnnoteFile{NoStop}{PhysRevLett.93.267201}%
\bibitem{PhysRevB.75.214515}%
  \BibitemOpen
  \bibfield{author}{%
  \bibinfo {author} {\bibfnamefont{E.}~\bibnamefont{Yusuf}}, \bibinfo {author}
  {\bibfnamefont{B.~J.}\ \bibnamefont{Powell}},\ and\ \bibinfo {author}
  {\bibfnamefont{R.~H.}\ \bibnamefont{McKenzie}},\ }%
  \bibfield{journal}{%
  \Doi{10.1103/PhysRevB.75.214515}{\bibinfo {journal} {Phys. Rev. B}}\ }%
  \textbf{\bibinfo {volume} {75}},\ \bibinfo {pages} {214515} (\bibinfo {month}
  {Jun}\ \bibinfo {year} {2007}),\
  \url{http://link.aps.org/doi/10.1103/PhysRevB.75.214515}%
  \bibAnnoteFile{NoStop}{PhysRevB.75.214515}%
\bibitem{Shiba01101975}%
  \BibitemOpen
  \bibfield{author}{%
  \bibinfo {author} {\bibfnamefont{H.}~\bibnamefont{Shiba}},\ }%
  \bibfield{journal}{%
  \Doi{10.1143/PTP.54.967}{\bibinfo {journal} {Progress of Theoretical
  Physics}}\ }%
  \textbf{\bibinfo {volume} {54}},\ \bibinfo {pages} {967} (\bibinfo {year}
  {1975}),\ \url{http://ptp.oxfordjournals.org/content/54/4/967.abstract}%
  \bibAnnoteFile{NoStop}{Shiba01101975}%
\bibitem{Hafer}%
  \BibitemOpen
  \bibfield{author}{%
  \bibinfo {author} {\bibfnamefont{H.}~\bibnamefont{H.~Hafermann}}, \bibinfo
  {author} {\bibfnamefont{P.}~\bibnamefont{Werner}},\ and\ \bibinfo {author}
  {\bibfnamefont{E.}~\bibnamefont{Gull}},\ }%
  \bibinfo {howpublished} {Computer Physics Communications, 184, 1280}
  (\bibinfo {year} {2013})%
  \bibAnnoteFile{NoStop}{Hafer}%
\bibitem{Bauer}%
  \BibitemOpen
  \bibfield{author}{%
  \bibinfo {author} {\bibfnamefont{B.}~\bibnamefont{Bauer}}, \bibinfo {author}
  {\bibfnamefont{L.~D.}\ \bibnamefont{Carr}}, \bibinfo {author}
  {\bibfnamefont{H.~G.}\ \bibnamefont{Evertz}}, \bibinfo {author}
  {\bibfnamefont{A.}~\bibnamefont{Feiguin}}, \bibinfo {author}
  {\bibfnamefont{J.}~\bibnamefont{Freire}}, \bibinfo {author}
  {\bibfnamefont{S.}~\bibnamefont{Fuchs}}, \bibinfo {author}
  {\bibfnamefont{L.}~\bibnamefont{Gamper}}, \bibinfo {author}
  {\bibfnamefont{J.}~\bibnamefont{Gukelberger}}, \bibinfo {author}
  {\bibfnamefont{E.}~\bibnamefont{Gull}}, \bibinfo {author}
  {\bibfnamefont{S.}~\bibnamefont{Guertler}}, \bibinfo {author}
  {\bibfnamefont{A.}~\bibnamefont{Hehn}}, \bibinfo {author}
  {\bibfnamefont{R.}~\bibnamefont{Igarashi}}, \bibinfo {author}
  {\bibfnamefont{S.~V.}\ \bibnamefont{Isakov}}, \bibinfo {author}
  {\bibfnamefont{D.}~\bibnamefont{Koop}}, \bibinfo {author}
  {\bibfnamefont{P.~N.}\ \bibnamefont{Ma}}, \bibinfo {author}
  {\bibfnamefont{P.}~\bibnamefont{Mates}}, \bibinfo {author}
  {\bibfnamefont{H.}~\bibnamefont{Matsuo}}, \bibinfo {author}
  {\bibfnamefont{O.}~\bibnamefont{Parcollet}}, \bibinfo {author}
  {\bibfnamefont{G.}~\bibnamefont{Pawłowski}}, \bibinfo {author}
  {\bibfnamefont{J.~D.}\ \bibnamefont{Picon}}, \bibinfo {author}
  {\bibfnamefont{L.}~\bibnamefont{Pollet}}, \bibinfo {author}
  {\bibfnamefont{E.}~\bibnamefont{Santos}}, \bibinfo {author}
  {\bibfnamefont{V.~W.}\ \bibnamefont{Scarola}}, \bibinfo {author}
  {\bibfnamefont{U.}~\bibnamefont{Schollwöck}}, \bibinfo {author}
  {\bibfnamefont{C.}~\bibnamefont{Silva}}, \bibinfo {author}
  {\bibfnamefont{B.}~\bibnamefont{Surer}}, \bibinfo {author}
  {\bibfnamefont{S.}~\bibnamefont{Todo}}, \bibinfo {author}
  {\bibfnamefont{S.}~\bibnamefont{Trebst}}, \bibinfo {author}
  {\bibfnamefont{M.}~\bibnamefont{Troyer}}, \bibinfo {author}
  {\bibfnamefont{M.~L.}\ \bibnamefont{Wall}}, \bibinfo {author}
  {\bibfnamefont{P.}~\bibnamefont{Werner}},\ and\ \bibinfo {author}
  {\bibfnamefont{S.}~\bibnamefont{Wessel}},\ }%
  \bibfield{journal}{%
  \Doi{10.1088/1742-5468/2011/05/P05001}{\bibinfo {journal} {Journal of
  Statistical Mechanics: Theory and Experiment}}\ }%
  \textbf{\bibinfo {volume} {2011}},\ \bibinfo {pages} {P05001} (\bibinfo
  {year} {2011}),\ \url{http://stacks.iop.org/1742-5468/2011/i=05/a=P05001}%
  \bibAnnoteFile{NoStop}{Bauer}%
\end{thebibliography}%

\end{document}